\documentstyle[]{article}
\def\appendix#1{
\addtocounter{section}{1}
\setcounter{equation}{0}
\renewcommand{\thesection}{\Alph{section}}
\section*{Appendix \thesection\protect\indent #1}
\addcontentsline{toc}{section}{Appendix \thesection\ \ \ #1}
}

\def\be{\begin{equation}}
\def\la{\label}
\def\ee{\end{equation}}
\def\bea{\begin{eqnarray}}
\def\eea{\end{eqnarray}}
\def\eps{\varepsilon}

\def\S{\sigma}
\def\n{\nabla}

\def\G{\Gamma}

\def\d{\delta}
\def\l{\left(}
\def\r{\right)}
\def\p{\partial}
\newcommand{\R}{{\rho }}
\newcommand{\M}{
{\mu}}
\newcommand{\N}{{\nu}}
\newcommand{\x}{\vec{x}}
\newcommand{\y}{\vec{y}}
\newcommand{\z}{\vec{z}}

\newcommand{\cE}{{\cal E }}
\begin{document}
\title{\hfill{UAHEP991} \\
\vspace{1cm}
Three-point Green function of the stress-energy tensor
in the AdS/CFT correspondence }
\author{
G.Arutyunov$^{a\, b}$\thanks{arut@genesis.mi.ras.ru}\mbox{}
 and \mbox{} S.Frolov$^{c\,b}$\thanks{frolov@bama.ua.edu}
\vspace{0.4cm} \mbox{} \\
$^a$ Universita di Milano,
\vspace{-0.1cm} \mbox{} \\
Dipartimento di Matematica "Federigo Enriques"
\vspace{-0.1cm} \mbox{} \\
Via C.Saldini, 50-20133, Milano, Italy.
\vspace{0.4cm} \mbox{} \\
$^b$Steklov Mathematical Institute,
\vspace{-0.1cm} \mbox{} \\
Gubkin str.8, GSP-1, 117966, Moscow, Russia
\vspace{0.4cm} \mbox{} \\
$^c$Department of Physics and Astronomy,
\vspace{-0.1cm} \mbox{} \\
University of Alabama, Box 870324,
\vspace{-0.1cm} \mbox{} \\
Tuscaloosa, Alabama 35487-0324, USA
\mbox{}
}
\date {}
\maketitle
\begin{abstract}
We compute the 3-point function of the stress-energy
tensor in the $d$-dimensional CFT from the $AdS_{d+1}$ gravity.
For $d=4$ the coefficients of the three linearly independent conformally
covariant forms entering the 3-point function are exactly the same as given by
the free field computations in the ${\cal N}=4$ SYM just as expected  from the
known renormalization theorems. For $d=3$ and $d=6$ our results give the
value of the corresponding 3-point function in the theories of strongly
coupled ${\cal N}=8$ superconformal scalar and $(2,0)$ tensor multiplets
respectively. 
\end{abstract}

\section{Introduction}
The Maldacena conjecture \cite{M}-\cite{W} relating
the large $N$ limit of certain conformal field theories
in $d$-dimensions with supergravity on the product of 
the $d+1$ dimensional Anti-de Sitter space with a 
compact manifold have been recently tested by explicit
computation of many 2- and 3-point gauge theory 
correlation functions from the AdS supergravity 
\cite{AV}-\cite{KR}. A considerable progress was also achieved in studying
4-point correlators \cite{LT1}-\cite{HF1}, whose complete computation, however,
requires the knowledge of the supergravity action on the AdS background
beyond the quadratic \cite{AF2} approximation.

 An important question being yet unsolved with 3-point functions
is computation of the 3-point function of the stress-energy
tensor $T_{ij}(\x )$. Unlike the other 3-point functions
that are determined by the conformal symmetry almost completely
(usually up to one constant), the 3-point function of $T_{ij}(\x )$
in general dimension $d$ admits five independent conformally 
covariant 
forms, two of them being fixed by the gauge theory
conservation law $\p_iT_{ij}(\x)=0$ \cite{Osb}. Three constants
undetermined neither conformal symmetry nor the conservation law
might be computed from the AdS supergravity and 
confronted with their free-field counterparts.
This obviously provides a further nontrivial test for 
the AdS/CFT correspondence.

In this paper we therefore address the problem of computing the 
3-point function of the stress-energy tensor in the 
$d$-dimensional CFT from
 the $AdS_{d+1}$ gravity.

According to the AdS/CFT conjecture the CFT current of conformal weight zero
coupled to the stress-energy tensor of conformal dimension $d$
is extended to the interior of the AdS space as the on-shell 
graviton field. In comparison with 3-point functions 
of other gauge-invariant composite operators
computation of the 3-point function of $T_{ij}(\x )$
is complicated by two things. Firstly to cure up the infrared divergences of
the gravity action one can introduce the boundary of the AdS space. The
Hamiltonian formulation \cite{AF} of the AdS/CFT correspondence then naturally
requires additional boundary terms \cite{LT}, so that the pure gravity action
is not given only by the standard Einstein-Hilbert term. It is worth
stressing that the account of the boundary terms is absolutely necessary
since they provide the fulfilment of the Ward identities in the boundary
CFT.  The second thing is that after introducing the boundary terms the
gravity action looses its manifest
 conformal as well as gauge
invariance. 

In our computation of the 3-point function we account the boundary terms in
the following way. We start with the standard  Einstein-Hilbert term (with the
cosmological constant) and decompose it up to the cubic order in metric
perturbation $h_{\mu\nu}$. Removing all terms linear in second derivatives as
well as all total derivative terms we are left with an action that according
to \cite{AF} differs from the action one should use in the AdS/CFT
correspondence only by terms that do not contribute to the Green functions.  
Then, by the on-shell perturbation theory, one may find that the contribution
of the quadratic terms to the value of the 3-point is exactly zero. Thus, the
remaining action is just the sum of the cubic bulk and the boundary
(noncovariant) terms, the latter arise due to the removal procedure.
Fortunately, the cubic boundary terms do not contribute to the
value of the 3-point function and by this reason can be disregarded.  The bulk
term is manifestly covariant w.r.t. the AdS isometries as well as on-shell
gauge symmetry and that apparently solves the second difficulty.

Choosing the covariant gauge we then explicitly compute 
the remaining bulk integral and get the 3-point function.
For the physically most interesting case $d=4$ we realize that the
coefficients of the conformal  tensors of the 3-point function are exactly
the same as the ones found by the free field  computations. This is obviously
in agreement with the earlier results by \cite{SSRS,HFS,HSW}, whose essence
is that in four dimensions the superconformal  symmetry is powerful enough to
protect some 3-point functions in ${\cal N}=4$ Yang-Mills against quantum
corrections. 

As to the other cases of particular interest $d=3$ and $d=6$, at the moment we
are unaware of the gauge theory calculations and may suggest that our results
give the value of the corresponding 3-point function in the theories of
the strongly coupled superconformal scalar and $(2,0)$ tensor multiplets
respectively.

The paper is organized as follows. In Section 2 we define the gravity action
one should use in the AdS/CFT correspondence  and obtain its decomposition up
to cubic order in metric  perturbation. In the third Section we compute the
on-shell value of the gravity action in the de Donder gauge and obtain the
3-point function. We then  write down explicitly the coefficients of the
conformal tensors occuring in the 3-point function in dimensions
$d=2,3,4,5,6$ and comment the  most
 interesting cases.
Some details of the calculation are collected in the Appendix.

\section{Gravity action at third order of perturbation} 
We begin by fixing the
basic notation. Let ${\cal M}$ be a  $d+1$ dimensional manifold with a
$d$-dimensional boundary.  Throughout the paper the indices $\mu,\nu, \ldots $
run the set $0, \ldots , d$, while $i,j,\ldots $ are 
reserved for the $d$-dimensional boundary and take 
values $1,\ldots, d$. The coordinates $x_{\mu}$ are
then splitted as $x_{\mu}=(x_0
,\x )$ with $\x=\{x_i\}$.
Let ${\cal M}$ be  also supplied with a positive signature
metric $G_{\mu\nu}$.

We will deal with the Euclidean version of the $AdS_{d+1}$
space that is described as the upper half space 
${\cal M}=\{x_i\in {\bf R};~~~x_0>0 \}$ endowed
with the metric 
$$
ds^2=\frac{1}{x_0^2}dx^{\mu}dx^{\mu}.
$$
The boundary of $AdS_{d+1}$ space is at $x_0=0$ and can be
identified with the Euclidean space. Since the boundary
is infinitely distant from any interior point the gravity
action on the AdS background suffers from infrared
divergences. A natural regularization is then provided 
by putting the boundary of $AdS_{d+1}$ space at $x_0=\eps$ and 
considering the part with $x_0\geq \eps $. The
physical fields are required to vanish when $x_0\to \infty$.

The total gravity action ${\bf S}$ one should use for computing 
gauge theory correlation functions is given by the sum
\bea
\la{tot}
{\bf S}=S+S^{(1)}+S^{(2)}
\eea
of the standard Einstein-Hilbert term ( with the cosmological 
constant $\lambda=\frac{1}{2}d(d-1)$)
\be
S=\int\sqrt{G}(R-2\lambda)
\la{EH}
\ee
and two boundary terms $S^{(1)}$ and $S^{(2)}$ \cite{LT}.
Explicitly, $S^{(1)}$ is the Gibbons-Hawking term \cite{GH}
\bea
\la{GH}
S^{(1)}=2\int_{\p M}d^{d}x \sqrt{\bar{G}}K,
\eea
where $K$ is the trace of the fundamental form on the boundary
and $\bar{G}$ is the determinant of the induced metric. 
The second boundary term $S^{(2)}$ is a term proportional to the 
volume of the boundary:
\bea
\la{BV}
S^{
(2)}=2(1-d)\int_{\p M}d^{d}x \sqrt{\bar{G}}.
\eea

As was shown in \cite{AF}, adding the boundary terms is equivalent 
to removing from the bulk action (\ref{EH}) 
all terms linear in second derivatives and all total derivative terms. 
The gravity action obtained in such a way differs from (\ref{tot}) 
only by terms which do not contribute to Green functions.  Therefore we can
restrict our attention to considering the bulk term. 

If we assume $g_{\M\N}$ to be the background AdS metric 
and perturb $G_{\M\N}$ near the background value:
$G_{\M\N}=g_{\M\N}+h_{\M\N}$, then 
equations of motion $R_{\M\N}=-dg_{\M\N}$ 
up to the second order in $h_{\M\N}$ can be written
as follows 
\bea
\la{em}
L_{\M\N}=V_{\M\N},
\eea
where two tensors
\bea
\la{L}
L_{\M\N}&=&\n^{\R}\n_{\R} h_{\M\N}+\n_{\M}\n_{\N}h-\n_{\M}
\n^{\R}h_{\N\R}-\n_{\N}\n^{\R}h_{\M\R}
+2(h_{\M\N}-g_{\M\N}h), 
\eea
and
\bea
\la{V}
V_{\M\N}&=&-\n_{\R}(h^{\R\S}(\n_{\M}h_{\N\S}+\n_{\N}h_{\M\S}
-\n_{\S}h_{\M\N}))
+\n_{\N}(h^{\R\S
}\n_{\M}h_{\R\S})\\
\nonumber
&+&
\frac{1}{2}(\n_{\M}h_{\N\R}+\n_{\N}h_{\M\R}-
\n_{\R}h_{\M\N})\n^{\R}h 
-\frac{1}{2}\n_{\M}h_{\R\S}\n_{\N}h^{\R\S}+
\n_{\R}h_{\M\S}\n^{\R}h_{\N}^{\S}
-\n_{\S}h_{\M\R}\n^{\R}h_{\N}^{\S}
\eea
were introduced.
Here the covariant derivatives are taken w.r.t the background metric.

Introduce the notation
$$
R_{\M\N}=\dot{R}_{\M\N}+R_{\M\N}^{(1)}+R_{\M\N}^{(2)}+...=
\dot{R}_{\M\N}+\delta R_{\M\N}+\frac{1}{2!}\d^{(2)}R_{\M\N}+...
$$ 
for decomposition of the Ricci 
tensor around the background
$\dot{R}_{\M\N}$ and analogous one for decomposition of the
curvature. 

Now we are ready to analyse gravity action (\ref{tot}) 
up to the third order in metric perturbation. We  
start with working out decomposition of (\ref{EH}):
\bea
\la{dec}
S=\dot{S}+\d S +\frac{1}{2!}\d^{(2)}S+\frac{1}{3!}\d^{(3)}S+...
\eea  
Computing the first variation $\d S$ of (\ref{EH}) one then 
represents it in the form
\bea
\la{fv}
\d S[G,h]=-\int_M \sqrt{G} \l R_{\M\N}-\frac{1}{2}G
_{\M\N}R+\lambda G_{\M\N}\r h^{\M\N}
+T,
\eea
where $T$ is the following boundary term 
$$
T=\int_M \sqrt{G}\n_{\M}(\n_{\N} h^{\M\N}-\n^{\M} h)=
-\int_{\p M}\sqrt{G}(\n_{\M}h^{0\M}-\p^0 h).
$$
In (\ref{fv}) we consider $\d S[G,h]$ as the variation of $S[G]$
at a "point" $G_{\M\N}$, i.e. we do not assume 
the metric $G_{\M\N}$ to 
be equal to its background value. Now the simple algorithm to
find decomposition (\ref{dec}) is to consider the
succesive variations of 
$\d S[G,h]$\footnote{The covariant derivatives
in $T$ are also w.r.t. to the metric $G_{\M\N}$.}. 
Since we are interested in  decomposition of the total action (\ref{tot}) we 
can omit the total derivative term $T$.

Thus, varying the terms in parenthesis in (\ref{fv}) and
reducing the result to the background we get 
$$
R^{(1)}_{\M\N}-\frac{1}{2}g^{(1)}_{\M\N}\dot{R}-\frac{1}{2}g_{\M\N}R^{(1)}
+\lambda g^{(1)}_{\M\N}=-\frac{1}{2}\l
L_{\M\N}-\frac{1}{2}g_{\M\N}L\r ,
$$
where the relation  
$R^{(1)}_{\M\N}+dh_{\M\N}=-\frac{1}
{2}L_{\M\N}$ was used.

For the second variation we find
$$
\d^{(2)}R_{\M\N}-\frac{1}{2}g_{\M\N}\d^{(2)}R
-\frac{1}{2}\d^{(1)}g_{\M\N}\d^{(1)}R=
2(R_{\M\N}^{(2)}-\frac{1}{2}g_{\M\N}(g^{\R\S}R_{\R\S}^{(2)}))-
\frac{1}{2}g_{\M\N}h^{\R\S}L_{\R\S}+\frac{1}{2}h_{\M\N}L.
$$
Here 
$R_{\M\N}^{(2)}-\frac{1}{2}g_{\M\N}(g^{\R\S}R_{\R\S}^{(2)})=
\frac{1}{2}(V_{\M\N}-\frac{1}{2}g_{\M\N}V)$, where $V=V_{\M}^{\M}$.

With these formulae at hand it is now easy to find 
the action (\ref{EH}) up to the third order in $h_{\M\N}$:
\bea
\la{ac}
S&=&\dot{S}+\int_M \sqrt{g}\left(
\frac{1}{4}(L_{\M\N}-\frac{1}{2}g_{\M\N}L)h^{\M\N}-\frac{1}{6}
(V_{\M\N}-\frac{1}{2}g_{\M\N}V)h^{\M\N} \right.\\
\nonumber
&-&\left.\frac{1}{6}\left(\frac{1}{4}h^2L-hh^{\M\N}L_{\M\N}
+2h^{\M\R}h_{\R}^{\N}L_{\M\N}-\frac{1}{2}h^{\M\N}h_{\M\N}L\right)\right) ,
\eea 
where we again omitted nonessential total derivative terms. 
The action (\ref{ac}) depends on second derivative terms.  To remove these 
terms one should add to (\ref{ac}) total derivative terms  which can be 
easily found by using explicit expressions (\ref{L}) and  (\ref{V}) for $L$ 
and $V$ respectively. A simple  consideration then shows that the quadratic 
terms in the resulting action do not contribute to the 3-point Green function. 

Thus,  we see
 that to find the 3-point function we need to compute  the on-shell value of 
${\bf S}$ which is given by \bea
\la{onsh}
{\bf S} &=&-\frac{1}{6}\int_M \sqrt{g}(V_{\M\N}-\frac{1}{2}g_{\M\N}V)h^{\M\N}
+ cubic ,
\eea
Here $h_{\M\N}$ is a solution to the linearized equation of motion:
$L_{\M\N}[h]=0$ and $cubic$ refers to the unwritten explicitly total derivative
terms of the cubic order. Since these terms may deliver only
a local contribution to the value of the 3-point function in what 
follows we disregard them.

\section{Three-point Green function}
The radiation 
gauge for the AdS gravity that is effectively used for computing
the two-point Green function of the stress-energy tensor in the boundary CFT
obviously breaks the invariance of the gravity action under isometries. For the
3-point function this fact leads to severe difficulties in computing the bulk
integrals. Thus, to handle the problem  we choose the covariant gauge of the
de Donder type:
\bea 
\la{Don}
\n_{\M}\left(h^{\M}_{\N}-\frac{1}{2}\d^{\M}_{\N} h\right)=0.
\eea 
In this gauge the solution 
of the linearized equations of motion
reads as  \cite{LT}:
\bea
\la{cp}
h_{\M}^{\N}(x_0,\vec{x})=\kappa_G \int d^dy~ 
{\cal K}(x,\vec{y})J_{\M}^i(x-\vec{y})J_j^{\N}(x-\vec{y})
{\cal E}_{ij,kl} h_{kl} (\vec{y}),
\eea
where 
$$
J_{\M}^{\N}(x)=\d_{\M}^{\N}-2\frac{x_{\M}x^{\N}}{|x|^2},~~~  
{\cal K}(x,\vec{y})=\frac{x_0^d}{(x_0^2+(\x-\y)^2)^d},  
$$
the coefficient $k_G=\frac{d+1}{d-1}\frac{\G(d)}{\pi^{d/2}\G(d/2)}$,
$h_l^k (\y)$  represents the boundary data of the graviton and 
${\cal E}_{ij,kl
}$ is the traceless symmetric projector:
\bea
\la{covsol}
{\cal E}_{ij,kl}=\frac{1}{2}(\d_{ik}\d_{jl}+\d_{il}\d_{kj})
-\frac{1}{d}\d_{ij}\d_{kl} .
\eea
Note that tensor $h_{\M}^{\N}$ has the vanishing trace.

In the de Donder gauge the remaining bulk term of 
the on-shell action 
(\ref{onsh}) can be represented in the following form
most suitable for further computations:
\bea
\nonumber
 {\bf S}&=&-\frac{1}{6}\int_{M} \sqrt{-g}\l V_{\M\N}-\frac{1}{2}g_{\M\N}V\r 
h^{\M\N} \\ \nonumber
&=&
\frac{1}{6}\int_{M} \sqrt{-g}\l
\n_{\S}(h^{\M\N}h_{\M\R}\n^{\R}h^{\S}_{\N})
-\frac{1}{2}\n_{\M}(h^{\M\N}h^{\R\S}\n_{\N}h_{\R\S})
-\n_{\S}(h_{\M}^{\S}h_{\N}^{\R}\n_{\R}h^{\M\N}) \r
\nonumber \\
\nonumber
%\la{fnm}
&-& \frac{1}{4}\int_{M}\sqrt{-g}\left(
\n_{\S}\n_{\R} h_{\M\N} h^{\M\N} h^{\R\S}  
-2\n_{\S}\n_{\R} h^{\M\N} h_{\M}^{\R} h^{\S}_{\N}
+\frac{2}{3}h_{\M}^{\N}\n_{\R}h_{\S}^{\M}\n^{\R}h_{\N}^{\S}
-\frac{2}{3}(d+1) h_{\M}^{\N} h^{\M}_{\S} h^{\S}_{\N}
\right),
\eea
where we used the explicit form of $V_{\M\N}$
and the vanishing of $h_{\M}^{\M}$.

Again omitting the total derivatives being
the cubic order boundary terms we see that
the on-shell action in the de Donder gauge is 
essentially given by the bulk integral
\bea
\la{f'}
{\bf S} = -\frac{1}{4}\int_{M}\sqrt{-g}\left(
\n_{\S}\n_{\R} h_{\M\N} h^{\M\N} h^{\R\S}  
-2\n_{\S}\n_{\R} h^{\M\N} h_{\M}^{\R} h^{\S}_{\N}
+\frac{2}{3}h_{\M}^{\N}\n_{\R}h_{\S}^{\M}\n^{\R}h_{\N}^{\S}
-\frac{2}{3}(d+1) h_{\M}^{\N} h^{\M}_{\S} h^{\S}_{\N}
\right).
\eea
By using the equation of motion that in the 
covariant gauge reads as $\n_{\R}\n^{\R}h_{\M\N}=-2h_{\M\N}$
we then rewrite (\ref{f'}) in the form
\bea
\la{f}
{\bf S} =\frac{1}{4}\int_{M}\sqrt{-g}\left(
2\n_{\S}\n_{\R} h^{\M\N} h_{\M}^{\R} h^{\S}_{\N}
-\n_{\S}\n_{\R} h_{\M\N} h^{\M\N} h^{\R\S} 
+\frac{2}{3}d h_{\M}^{\N} h^{\M}_{\S} h^{\S}_{\N}
\right).
\eea

The computation of (\ref{f}) is a 
rather combersome but
purely technical task that can be performed by the inversion 
method 
of \cite{FMMR}. Before plugging into
details  we make some comments 
about the relation between the bulk and the boundary gauge 
transformations. The symmetry group of action (\ref{f}) is 
now reduced to the gauge transformations that preserve
the de Donder gauge. This group of residual gauge 
transformations is generated by vectors $\xi^{\M}$ obeying
an equation $\n_{\R}\n^{\R}\xi_{\M}-d\xi_{\M}=0$. Explicitly,
the solution satisfying the gauge $\n_{\M}\xi^{\M}=0$
reads as 
\bea
\la{cv}
 \xi^{\M}(x_0,\x)=\kappa_v\int d^dy 
\frac{x_0^{d+2}}{(x_0^2+(\x-\y)^2)^{d+1}}J_{i}^{\mu}(x-\y)\xi^i(\y),
~~~~\kappa_v=\frac{d+1}{d}\frac{2\G(d)}{\pi^{d/2}\G(d/2)},
\eea
where the coefficient $\kappa_v$ is fixed by requiring
$\xi^{i}(x_0,\x)\to \xi^{i}(\x)$ when $x_0\to 0$. 
In particular, the component $\xi^0$ is 
$$
\xi^0(x_0,\x)=\frac{\kappa_v}{d+1}
\int d^dy \frac{x_0^{d+3}}{(x_0^2+(\x-\y)^2)^{d+1}}\p_i\xi^i(\y).
$$
One may see that on the boundary $(\eps\to 0)$ the residual
gauge transformations are 
reduced to 
\bea
\la{usu}
\d h_i^j=\p_i\xi^j+\p_j\xi^i-\frac{2}{d}\d_{ij}(\p_k\xi^k),
\eea
i.e. to the usual gauge transformations of a traceless symmetric
tensor. In other words, the transformations (\ref{usu}) of the
boundary data can always be prolonged to the bulk gauge fields,
which preserve the de Donder gauge. We, therefore, expect
the 3-point function 
$T_{ij,kl,mn}(\x,\y,\z)=
<T_{ij}(\x )T_{kl}(\y )T_{mn}(\z )>$
of the stress-energy tensor $T_{ij}$ to obey the 
conservation law 
$$
\p_i T_{ij,kl,mn}(\x, \y ,\z )=0,~~~
\mbox{for noncoincident $\x , \y $ and $\z$.}
$$

As was already mentioned in the Introduction
in arbitrary dimension $d>3$
there are five independent conformal tensors occuring
in the expression for the 3-point function of the 
stress-energy tensor. The conservation law then
fixes the value of two from five coefficients. It is clear
that the 3-point function defined
by $S$ is conformally covariant and the only reason to find
its explicit expression is to make
 comparison of  the 
coefficients of conformal tensors to the ones found 
on the gauge theory side.

Thus, substituting (\ref{cp}) in (\ref{f}) we see that 
according to the AdS/CFT prescription \cite{GKP,W}
the 3-point function is defined as\footnote{We assume that
the coupling of $T_{ij}(\x)$ with $h_{ij}(\x)$ on the 
boundary of $AdS_{d+1}$ is given by 
$\int d^dx\l\frac{1}{2}T_{ij}(\x)h_{ij}(\x)\r $
and this explain the number $8$ in (\ref{sym}).
Later on we show that this coupling also leads
 to the 
correct Ward identity.}
\bea
\la{sym}
T_{ij,kl,mn}(\x,\y,\z)=8\sum I_{ij,kl,mn}(\x ,\y ,\z ),
\eea
where sum is taken over all possible permutations of 
sets of indicies 
and points $(ij, \x )$, $(kl, \y )$ and $(mn, \z )$
of the following tensor
\bea
\la{f1}
I_{ij,kl,mn}(\x,\y,\z)
&=&2\kappa_G^3 
{\cal E}_{ij,i'j'}{\cal E}_{kl,k'l'}{\cal E}_{mn,m'n'}
\int \frac{d^{d+1}\omega}{\omega_0^{d+1}}
{\cal K}(\omega,\y ){\cal K}(\omega,\z )\times \\
\nonumber
&& \left[ 
%%%%%%%%%%%%%%%%%
%%%%%%%%%%%%%%%%%%%%%%%%%%%%%%%%%%%%%%%%
2\n^{\S}\n_{\R}({\cal K}J^{\M}_{i'}J_{\N}^{j'})(\omega-\x )
(J_{\M}^{k'}J^{\R}_{l'})(\omega-\y ) 
(J_{\S}^{m'}J_{n'}^{\N})(\omega-\z ) \right. \\
%%%%%%%%%%%%%%%%%%%%%%%%%%%%%%%%%%%%%%%%%%%%%%%%%%%%%%
\nonumber
&-&\n_{\S}\n^{\R}({\cal K}J^{\M}_{i'}J_{\N}^{j'})(\omega-\x) 
(J_{\M}^{k'}J^{\N}_{l'})(\omega-\y)
(J_{\R}^{m'}J^{\S}_{n'})(\omega-\z) \\
%%%%%%%%%%%%%%%%%%%%%%%%%%%%%%%%%%%%%%%%%%%%%%%%%%%%%%%%%%
\nonumber
&+&\frac{2d}{3} \left.
({\cal K}J^{\M}_{i'
}J_{\N}^{j'})(\omega-\x )
(J_{\M}^{k'}J^{\R}_{l'})(\omega-\y ) 
(J_{\R}^{m'}J_{n'}^{\N})(\omega-\z )
\right]. 
\eea
Recall that here and in what follows the bulk (boundary) indices 
are contracted w.r.t. the AdS metric (Euclidean) and, therefore,
only their positions matter. Note that the tensor $I_{ij,kl,mn}$
itself is not conformally covariant.

Following the method of \cite{FMMR} we now 
put in (\ref{f1}) $\x =0$ and perform 
the change of variables
$\omega'_{\M}=\frac{\omega_{\M}}{\omega^2}
$ and $x'_i=\frac{x_i}{x^2}$.
This is just the inversion transformation under which
the derivatives $\n_{\M}$ transform covariantly.
In particular,
$$
\n_{\R}\l {\cal K}(\omega ,\x)J_{\M}^i(\omega -\x )
J_j^{\N}(\omega -\x )\r
=|\omega'|^2 J_{\M}^{\lambda}(\omega')J_{\beta}^{\N}(\omega')
J_{\R}^{\S}(\omega')\n_{\S}'
\l {\cal K}(\omega' ,\x ')J_{\lambda}^{a}(\omega' -\x')
J_{b}^{\beta}(\omega' -\x')\r
\frac{J_{a}^{i}(\x)J_{j}^{b}(\x)}{|x|^{2d}}
$$ 
and 
$$
\n_{\R}\l {\cal K}(\omega ,0)J_{\M}^i
(\omega )
J_j^{\N}(\omega )\r
=|\omega'|^2 J_{\M}^{\lambda}(\omega')J_{\beta}^{\N}(\omega')
J_{\R}^{\S}(\omega')\n_{\S}'
\l (\omega_0')^d \d_{\lambda}^{i}
\d_{j}^{\beta}\r , 
$$ 
where the covariant derivatives $\n_{\M}' $ are w.r.t to the 
connection
$$
\Gamma'^\M_{\R\S}(\omega')=-\frac{1}{{\omega_0}'}
(\d_{\R}^0\d_{\M\S}+\d_{\S}^0\d_{\M\R}-\d_{\M}^0\d_{\R\S}).
$$
Thus, after substituting the change of variables all
internal Jacobians
depending on the variable $\omega$ alone cancel against 
each other and one is left with the following expression
\bea
\la{f3}
&& 
I_{ij,kl,mn}(0,\y ,\z )
= 2\kappa_G^3 
{\cal E}_{ij,i'j'}
\frac{{\cal I}_{kl,k'l'}(\y) }{|y|^{2d}}
\frac{{\cal I}_{mn,m'n'} (\z) }{|z|^{2d}}
\int \frac{d^{d+1}\omega }{\omega_0}
{\cal K}(\omega,\y'){\cal K}(\omega,\z')\times \\
\nonumber
&&
\left( -\frac{4d}{3}(J_{i'}^{k'} J_{l'}^{s})(\omega-\y')(J_{s}^{m'} 
J_{n'}^{j'})(\omega-\z')
 +2(d^2+\frac{d}{3}+2)(J_{i'}^{k'} J_{l'}^0)(\omega-\y')(J_{j'}^{m'} 
J_{n'}^0)(\omega-\z' )
\right. \\
\nonumber
&& \left.
-(d^2-d-2)(J_{i'}^{k'} J_{l'}^{j'})(\omega-\y')
(J_{0}^{m'} J_{n'}^0)(\omega-\z') 
+2d(J_{0}^{k'} J_{l'}^0)(\omega-\y')(J_{i'}^{m'} J_{n'}^{j'})(\omega-\z')
\right),
\eea
where the concise notation 
${\cal I}_{ij,kl}(\x)=\cE_{ij,i'j'}(J_{i'k}J_{j'l})(\x)$
was introduced.

In view of (\ref{sym})
it is further more convenient to deal with the integral 
$I^{sym}_{ij,kl,mn}$ being the symmetrization of (\ref{f3})
w.r.t. $(kl,\y )$ and $(mn,\z )$: 
\bea
\la{Is}
I^{sym}_{ij,kl,mn}(0,\y,\z)=I_{ij,kl,mn}(0,\y,\z)+I_{ij,mn,kl}(0,\z,\y).
\eea

The computation of $I^{sym}$ is sketched in the Appendix and
below we present the result
\bea
\la{Gen}
&&I^{sym}_{ij,kl,mn}(0,\y,\z)=
2\pi^{d/2}\kappa_G^3\frac{\G(d/2)(\G(d/2+1))^2}{(d+1)^2\G(d-1)\G(d+2)}
{\cal E}_{ij,i'j'}
\frac{{\cal I}_{kl,k'l'}(\y) }{|y|^{2d}}
\frac{{\cal I}_{mn,m'n'} (\z) }{|z|^{2d}}
\frac{1}{|t|^d}\times \\
\nonumber
&&\left[  
a_1 \cE_{i'j', ab}\cE_{k'l',ac}\cE_{m'n',bc} 
+a_2 \cE_{i'j', a
b}\cE_{k'l',ac}\cE_{m'n',bd}
\frac{t_{c}t_{d}}{t^2} \right. \\
\nonumber
&&+
a_3(\cE_{k'l', ab}\cE_{m'n',bc}\cE_{i'j',ad}+
(k'l')\to (m'n') )\frac{t_{c}t_{d}}{t^2}
+a_4\l \cE_{k'l', i'j'}
\l \frac{t_{m'}t_{n'}}{t^2}-\frac{1}{d}\d_{m'n'}\r +(k'l')\to (m'n')  \r  \\
\nonumber
&&+
a_5 \cE_{k'l',m'n'}\l \frac{t_{i'}t_{j'}}{t^2}-\frac{1}{d}\d_{i'j'}\r  
+a_6 
\l \cE_{k'l',ab}\cE_{i'j',ac}
\l \frac{t_{m'}t_{n'}}{t^2}-\frac{1}{d}\d_{m'n'}\r + (k'l')\to (m'n')\r
\frac{t_{b}t_{c}}{t^2}  \\
\nonumber
&&+
a_7\cE_{k'l',ab}\cE_{m'n',ac}\l
\frac{t_{i'}t_{j'}}{t^2}-\frac{1}{d}\d_{i'j'}\r \frac{t_{b}t_{c}}{t^2}  
+a_8 \l \frac{t_{i'}t_{j'}}{t^2}-\frac{1}{d}\d_{i'j'}\r
 \l \frac{t_{k'}t_{l'}}{t^2}-\frac{1}{d}\d_{k'l'}\r
 \l \frac{t_{m'}t_{n'}}{t^2}-\frac{1}{d}\d_{m'n'}
 \r  \left.\right].
\eea
In the last formula the Latin indices $(a,b,c,d)$ are used to distinguish
boundary summation indices and the variable 
$t_i=z_i'-y_i'$.
The coefficients $a_i$, $i=1,\ldots , 8$ are expressed through 
the constants $a_i^{(k)}$, $k=1,2,3$ found in the Appendix
by the following formula
\bea
\la{coef}
a_i=-\frac{4d}{3}a_i^{(1)}+2(d^2+\frac{d}{3}+2)a_i^{(2)}-(d^2-3d-2)a_i^{(3)}.
\eea

If we now restore 
the $\x$-dependence, the variable $t_i$:
$$
t_i=(z-x)_i'-(y-x)_i'
=\frac{(x-y)_i}{(\x-\y)^2}-\frac{(x-z)_i}{(\x-\z)^2}
$$ 
turns
into the conformal vector $X_i$: $t_i=-X_i$ with a remarkable
property to transform covariantly $X_i\to X_i'$ under the 
inversion $x_i'=\frac{x_i}{x^2}$ \cite{Osb}: 
$$
X'_i=x^2 J_{ij}(\x)X_j.
$$

Then by using the following two identities 
$$
J_{ij}(\x-\z)Z_{j}=-\frac{(\x-\y)^2}{(\z-\y)^2}X_{j},
~~~ 
J_{ik}(\x-\z)J_{kj}(\z-\y)=J_{ij}(\x-\y)+2(\x-\y)^2X_{i}Y_{j},
$$
one may finally represent the 3-point
function (\ref{sym}) in the form
\bea 
\nonumber
T_{ij,kl,mn}&=&\frac{1}{|\x-\y|^{2d}|\y-\z|^{2d}|\x-\z|^{2d}}
\left[
\cE_{ij, i'j'}\cE_{kl,k'l'}\cE_{mn,m'n'}
({\cal A} J_{i'k'}(\x-\y)J_{l'm'}(\y-\z)J_{j'n'}(\z-\x) \right. \\
\nonumber
&+& {\cal B}J_{i'k'
}(\x-\y)J_{j'n'}(\x-\z)Y_{l'}Y_{m'}(\y-\z)^2
+cycl.perm. ) \\
\la{Symmetr}
&+&{\cal C}\l {\cal I}_{ij,kl}\l \frac{Z_nZ_m}{Z^2}-\frac{1}{d}\d_{mn}\r
+cycl.perm.\r \\
\nonumber
&+&{\cal D}\l \cE_{ij,i'j'}\cE_{kl, k'l'}X_{i'}Y_{k'}(\x-\y)^2
J_{j'l'}(\x-\y)\l \frac{Z_mZ_n}{Z^2}-\frac{1}{d}\d_{mn}\r
+cycl.perm.\r  \\
\nonumber
&+&
\left.
{\cal E}
\l \frac{X_iX_j}{X^2}-\frac{1}{d}\d_{ij}\r
 \l \frac{Y_kY_l}{Y^2}-\frac{1}{d}\d_{kl}\r
 \l \frac{Z_mZ_n}{Z^2}-\frac{1}{d}\d_{mn} \r \right], 
\eea 
where 
\bea
\nonumber
{\cal A}=3\Delta_da_1,~~~{\cal B}=\Delta_d(2a_1+a_2-2a_3),~~~
{\cal C}=\Delta_d(2a_4+a_5), \\
\la{finc}
{\cal D}=\Delta_d(4a_5+a_7-4a_3-2a_6), 
~~~{\cal E}=\Delta_d(12a_5+6a_7+3a_8)
\eea
and 
$$
\Delta_d =
2\pi^{d/2}\kappa_G^3
\frac{\G(d/2)(\G(d/2+1))^2}{(d+1)^2\G(d-1)\G(d+2)}=
\frac{d\G(d)}{2\pi^d (d-1)^2}.
$$
As was expected formula (\ref{Symmetr}) is just the conformally
covariant 3-point function of the stress-energy 
tensor in the $d$-dimensional conformal field theory 
and it involves five independent conformal tensors. 

The coefficients ${\cal A},...,{\cal E}$ computed for 
any dimension $d\geq 2$ represent our basic result.
We first discuss the most interesting case $d=4$.
In this case by using the Table 1 of the Appendix 
one finds for ${\cal A},...,{\cal E}$ the 
following values

\bea 
\la{d4c} 
{\cal A}=-\frac{8\cdot 128}{9\pi^4},~~~
{\cal B}=-\frac{8\cdot 392}{9\pi^4},~~~ 
{\cal C}=-\frac{8\cdot 184}{9\pi^4},~~~
{\cal D}=-\frac{8\cdot 472}{9\pi
^4},~~~
{\cal E}=-\frac{8\cdot 304}{9\pi^4}. 
\eea 

Now we are ready to confront the coefficients (\ref{d4c}) with the ones found
by the free field computations in ${\cal N}=4$ SYM. To this end one needs to
restore the gravity coupling $g_G^{-2}$ that enters as an overall
constant in front of the total action ${\bf S}$. The value of the 
coupling 
constant is fixed by the type IIB supergravity on the $AdS_5\times S^5$ 
background and is equal to $$
g_G^2=\frac{8\pi^2}{N^2}.
$$
Clearly, to restore 
the $g_G^2$-dependence of  the 
3-point function 
we should multiply (\ref{d4c}) on $g_G^{-2}$ and obtain
\bea
\la{d4c1}
{\cal A}=-\frac{128}{9\pi^6}N^2,~~~
{\cal B}=-\frac{392}{9\pi^6}N^2,~~~
{\cal C}=-\frac{184}{9\pi^6}N^2.
\eea
Recall that the 2-point function of $T_{ij}(\x)$ found
from the $AdS_{d+1}$ gravity is given by 
\bea
\la{2p}
<T_{ij}(\x),T_{kl}(\y)>=\frac{C_d}{|\x-\y|^d}{\cal I}_{ij,kl}(\x-\y)
\eea
with the central charge $C_d=\frac{2k_G d}{g_G^2}$. 
 In particular, for $d=4$, one gets $C_4=\frac{80}{\pi^2 
g_G^2}=\frac{10}{\pi^4}N^2$.

In \cite{OP} it was shown that for any four-dimensional 
free field theory
given by $n_s$ scalars, by $n_f$ Dirac fermions and by $n_v$
vector fields the coefficients ${\cal A},{\cal B},{\cal C}$
are as follows 
\bea
\nonumber
&&{\cal A}=\frac{1}{\pi^6}\l \frac{8}{27}n_s-16n_v\r \\
\nonumber
&&{\cal B}=-\frac{1}{\pi^6}\l \frac{16}{27}n_s+4n_f+32n_v\r \\
\nonumber
&&{\cal C}=-\frac{1}{\pi^6}\l \frac{2}{27}n_s+2n_f+16n_v\r .
\eea
Substituting here the field-theoretical content of the 
${\cal N}=4$ $SU(N)$ SYM:
$$
n_s=6(N^2-1),~~~n_f=2(N^2-1),~~~n_v=N^2-1,
$$
we, thus, arrive at
\bea
\la{ftc}
{\cal A}=-\frac{128}{9\pi^6}(N^2-1),~~~
{\cal B}=-\frac{392}{9\pi^6}(N^2-1),~~~
{\cal C}=-\frac{184}{9\pi^6}(N^2-1).
\eea
The central charge $C_4$ can be found by 
taking into account the Ward identity that relates
$C_4$ with coefficients ${\cal A},{\cal B},{\cal C}$ of the 3-point 
function \cite{Osb}: $$
C_4=\frac{\pi^2}{12}(9{\cal A}-
{\cal B}-10{\cal C})=\frac{10}{\pi^4}(N^2-1).
$$
It is now obvious that in the large $N$ limit the coefficients  
${\cal A},{\cal B},{\cal C}$ and $C_4$ of the ${\cal N}=4$ $SU(N)$ SYM 
coincide with the ones found from the  $AdS_5$ gravity.

Thus, the coefficients $\cal A$, $\cal B$ and $\cal C$
of the conformal tensors obtained from
the $AdS_5$ gravity and reflecting thereby 
the strong-coupling behavior of the corresponding gauge theory
do not receive corrections to their free field (one-loop) values.
This fact finds a good agreement with the results by \cite{SSRS,HFS,HSW}.
Indeed, the traceless conserved stress-energy tensor
occurs in the multiplet of the supercurrent $T=tr(W^2)$,
where $W$ is an analytic superfield describing the ${\cal N}=4$
Yang-Mills strength multiplet. In \cite{SSRS} it was checked
for the leading components of $T$ being the scalar fields
that their 3-point functions computed from $AdS_5\times S^5$
supergravity coincide with the one-loop results in the large
$N$ limit. The same conclusion about vanishing of the radiative corrections 
at order $g^2$ was achieved in \cite{HFS} even for finite $N$.
By considering the anomaly in the superconformal  symmetry, it has been argued
in \cite{HSW} (see also \cite{FMMR})  that the 2- and 3-point functions of $T$
should actually  have a one-loop nature.

Having discussed the four-dimensional case, 
we now list explicitly the coefficients ${\cal A},...,{\cal E}$ for 
dimensions $d=2,3,5,6$ that can be found from (\ref{coef})
 together    with
(\ref{a1})-(\ref{a3}).

\underline{$d=2$}:   
\bea  
\la{d2} 
{\cal A}=-\frac{32}{\pi^2},~~~
{\cal B}=-\frac{40}{\pi^2}, ~~~
{\cal C}=-\frac{12}{\pi^2},~~~ 
{\cal D}=-\frac{32}{\pi^2},~~~
{\cal E}=-\frac{16}{\pi^2}.     
\eea

\underline{$d=3$}:
\bea  
\la{d3}
{\cal A}=-\frac{3^4}{2\pi^3},~~~ 
{\cal B}=-\frac{19\cdot 3^2}{2\pi^3},~~~
{\cal C}=-\frac{11\cdot 3^3}{2^3\pi^3},~~~ 
{\cal D}=-\frac{41\cdot 3^2}{2^2\pi^3},~~~
{\cal E}=-\frac{11\cdot 3^4}{2^4\pi^3}.  
\eea

\underline{$d=5$}:  
\bea 
\la{d5}
{\cal A}=-\frac{3\cdot 5^4}{2^2 \pi^5},~~~  
{\cal B}=-\frac{303\cdot 5^2}{2^2\pi^5},~~~ 
{\cal C}=-\frac{117\cdot 5^3}{2^4\pi^5},~~~ 
{\cal D}=-\frac{257\cdot 3 \cdot 5^2}{2^3\pi^5},~~~ 
{\cal E}=-\frac{137\cdot 3\cdot 5^3}{2^5\pi^5}. 
\eea

\underline{$d=6$}: 
\bea  
\la{d6}
{\cal A}=-\frac{2^8 \cdot 3^5}{5^2\pi^6},~~~ 
{\cal B}=-\frac{181\cdot 2^6 \cdot 3^3}{5^2\pi^6},~~~    
{\cal C}=-\frac{59\cdot 2^5 \cdot 3^4}{5^2\pi^6},~~~  
{\cal D}=-\frac{119\cdot
 2^7\cdot 3^3}{5^2\pi^6},~~~ 
{\cal E}=-\frac{2^7 \cdot 3^7}{5^2\pi^6}.   
\eea

In \cite{Osb} it was shown that the conservation law  implies the fulfilment
of the  following two identities  
\bea
\nonumber
&&(d^2-4){\cal A}+(d+2){\cal B}-4d{\cal C}-2{\cal D}=0,\\
\nonumber
&&(d-2)(d+4){\cal B}-2d(d+2){\cal C}+8{\cal D}-4{\cal E}=0.
\eea
It is needless to say that coefficients (\ref{d4c}) and (\ref{d2})-(\ref{d6})
satisfy both of them. It is, of course, only the check that we have done the
computation of (\ref{f}) correctly. One should be also aware of the fact that
for $d=2$ and $d=3$ the number of linearly independent conformal tensors is
reduced to 1 and 2 respectively \cite{OP}.

The cases $d=3$ and $d=6$ are of particular interest since according to the
AdS/CFT conjecture they correspond 
to compactifications of the 11d supergravity on $AdS_4\times S^{7}$ and
$AdS_7\times S^{4}$ respectively.
We, therefore, expect that coefficients (\ref{d3}) describe the 3-point
function of $T_{i
j}(\x)$ in the strongly coupled three-dimensional gauge
theory of the ${\cal N}=8$ conformal scalar multiplet while (\ref{d6}) are
expected to be found in the six-dimensional gauge theory of the $(0,2)$
tensor multiplet \cite{S}-\cite{CKvP}.

 \vskip 1cm
{\bf ACKNOWLEDGMENT} G.A. would like to thank Prof.F.Magri
for the kind hospitality at the Dipartimento di Matematica dell' 
Universita di Milano, where this work was completed.  The work of G.A. was
supported by the Cariplo Foundation for Scientific Research and in part by the
RFBI grant N96-01-00608, and the work of S.F. was supported by
the U.S. Department of Energy under grant No. DE-FG02-96ER40967
and in part by the RFBI grant N96-01-00551.
%%%%%%%%%%%%%%%%%%%%%%%%%%%%%%%%%%%%%%%%%%%%%%%%%%%%%%%%%%%%%%%%

\setcounter{section}{0}
\appendix{}
\setcounter{equation}{0}
We sketch here for reader's convenience some details 
of computation of integral (\ref{Is}).

Clearly, (\ref{Is}) can be written as the following sum 
of the integrals
 $I^k$, $k=1,2,3$:
\bea
%\la{I1sym}
\nonumber
I^{sym}_{ij,kl,mn}&=& = \frac{\kappa_G^3}{2}
{\cal E}_{ij,i'j'}
\frac{{\cal I}_{kl,k'l'}(\y) }{|y|^{2d}}
\frac{{\cal I}_{mn,m'n'} (\z) }{|z|^{2d}} \times \\
\nonumber
&&\left(
-\frac{4d}{3}I^{1}_{i'j',k'l',m'n'}
+2(d^2+\frac{d}{3}+2)I^{2}_{i'j',k'l',m'n'}
-(d^2-d-2)I^{3}_{i'j',k'l',m'n'}
\right),
\eea
where $I^k$ are given by
\bea 
%\la{I1}
\nonumber
I^1_{ij,kl,mn}(0,\y ,\z )
={\cal E}_{ij,i'j'}{\cal E}_{kl,k'l'}{\cal E}_{mn,m'n'}
\int \frac
{d^{d+1}\omega}{\omega_0}
{\cal K}(\omega,\y'){\cal K}(\omega,\z')
(J_{i'}^{k'} J_{l'}^{s})(\omega-\y')
(J_{s}^{m'} J_{n'}^{j'})(\omega-\z'),
\eea
\bea
%\la{I2}
\nonumber
I^2_{ij,kl,mn}(0,\y ,\z )
={\cal E}_{ij,i'j'}{\cal E}_{kl,k'l'}{\cal E}_{mn,m'n'}
\int \frac{d^{d+1}\omega}{\omega_0}
{\cal K}(\omega ,\y'){\cal K}(\omega,\z')
(J_{i'}^{k'} J_{l'}^0)(\omega-\y')
(J_{j'}^{m'} J_{n'}^0)(\omega-\z')
\eea
\bea
%\la{I3}
\nonumber
I^3_{ij,kl,mn}(0,\y,\z )
&&={\cal E}_{ij,i'j'}{\cal E}_{kl,k'l'
}{\cal E}_{mn,m'n'}
\int \frac{d^{d+1}\omega}{\omega_0}
{\cal K}(\omega ,\y'){\cal K}(\omega,\z')\times \\
\nonumber
&&\frac{1}{2}
\l
(J_{j'}^{k'} J_{l'}^{i'})(\omega-\y') 
(J_{0}^{m'} J_{n'}^0)(\omega-\z') 
+(J_{0}^{k'} J_{l'}^0)(\omega-\y') 
(J_{j'}^{m'} J_{n'}^{i'})(\omega-\z') 
\r .
\eea 

Then by using the identities
%%%%%%%%%%%%%%%%%%%%%%%%
\bea
\nonumber
&&\cE_{kl,k'l'}(J_{i}^{k'}J_{l'}^{j})(\omega)
\frac{\omega_0^d}{(\omega_0^2+\vec{\omega}^2)^d}=\\
\nonumber
&&\cE_{ij,k'l'}\left
( \frac{d-1}{d+1}
\d_i^{k'}\d^j_{l'}\frac{\omega_0^{d}}{(\omega_0^2+\vec{\omega}^2)^d}
-\frac{1}{2d(d+1)}\p_{k'}\p_{i}
\frac{\d^{j}_{l'}\omega_0^{d}}{(\omega_0^2+\vec{\omega}^2)^{d-1}}
-\frac{1}{2d(d+1)}\p_{l'}\p_{j}
\frac{\d_{i}^{k'}\omega_0^{d}}{(\omega_0^2+\vec{\omega}^2)^{d-1}}
\right.  \\
\nonumber
&& +
\left.
\frac{1}{2(d-1)d(d+1)}\p_{k'}\p_{l'}
\frac{\d_{i}^{j}\omega_0^{d}}{(\omega_0^2+\vec{\omega}^2)^{d-1}}
+\frac{1}{4(d-2)(d-1)d(d+1)}\p_i\p_j\p_{k'}\p_{l'}
\frac{\omega_0^{d}}{(\omega_0
^2+\vec{\omega}^2)^{d-2}}\right),
\eea
%%%%%%%%%%%%%%%%%%%%%%%%%%%%%%%%%%%%%%%%%%%%%%
\bea
\nonumber
&&\cE_{ij,i'j'}\cE_{kl,k'l'}(J_{i'}^{k'}J_{l'}^{j'})(\omega)
\frac{\omega_0^d}{(\omega_0^2+\vec{\omega}^2)^d}
=\cE_{ij,i'j'}\cE_{kl,k'l'} \\
\nonumber
&&\l\frac{d-1}{d+1}\d_{i'}^{k'}\d_{l'}^{j'}
\frac{\omega_0^d}{(\omega_0^2+\vec{\omega}^2)^d}
-\frac{1}{d(d+1)}\p_{k'}\p_{j'}
\frac{\d_{l'}^{i'}\omega_0^d}{(\omega_0^2+\vec{\omega}^2)^{d-1}}
+\frac{1}{4(d-2)(d-1)d(d+1)}\p_{i'}\p_{j'}\p_{k'}\p_{l'
}
\frac{\omega_0^d}{(\omega_0^2+\vec{\omega}^2)^{d-2}}
\r ,
\eea
%%%%%%%%%%%%%%%%%%%%%%%%%%%%%%%%%%%%%%%%%%%%%%%%%%%%%%
\bea
\nonumber
&&\cE_{kl,k'l'}(J_{i}^{k'}J_{l'}^{0})(\omega)
\frac{\omega_0^d}{(\omega_0^2+\vec{\omega}^2)^d}\\
\nonumber
&&=\frac{d-1}{d(d+1)}
\cE_{kl,im}\p_m
\frac{\omega_0^{d+1}}{(\omega_0^2+\vec{\omega}^2)^d}
-\frac{1}{2(d-1)d(d+1)}{\cal E}_{kl,k'l'}
\p_{k'}\p_{l'}\p_{i}
\frac{\omega_0^{d+1}}{(\omega_0^2+\vec{\omega}^2)^{d-1}}
\eea
%%%%%%%%%%%%%%%%%%%%%%%%%%%%%%%%%%%%%
%%%%%%%%%%%%%%%%%%%%%%%%%
and
\bea 
\nonumber
\cE_{ij,i'j'}(J_{i'}^{0}J_{0}^{j'})(\omega)
\frac{\omega_0^d}{(\omega_0^2+\vec{\omega}^2)^d}
=\frac{1}{d(d+1)}\cE_{ij,i'j'}\p_{i'}\p_{j'}
\frac{\omega_0^{d+2}}{(\omega_0^2+\vec{\omega}^2)^d} .
\eea
one can rewrite every $I^{k}$ as derivatives with respect
to the external variable $t_i=z'_i-y'_i$ of the standard 
integrals \cite{FMMR}:
\bea
\nonumber
&&I^a_{b,c}=\int d^{d+1}\omega \frac{\omega_0^a}
{[\omega_0^2+\vec{\omega}^2]^b [\omega_0^2+(\vec{\omega}-\vec{t})^2]^c} 
=\frac{\pi^{d/2}}{2}
\frac{\G(a/2+1/2)\G(b+c-d/2-a/2-1/2)}{\G(b)\G(c)} \times \\
\nonumber
&&
\frac{\G(1/2+a/2+d/2-b)\G(1/2+a/2+d/2-c)}
{\G(1+a+d-b-c)}|\x-\y |^{1+a+d-2b-2c}.
\eea

After straightforward calculations one obtains that 
all integrals $I^{k}$ result in the same form
\bea
%\la{Geni}
\nonumber
I^{k}&=&
\frac{\pi^{d/2}}{2}\frac{\G(d/2)(\G(d/2+1))^2}{(d+1)^2\G(d-1)\G(d+2)}
\frac{1}{|t|^d} \left[ \right.\\
\nonumber
&+& 
a_1^{(k)} \cE_{ij, ab}\cE_{kl,ac}\cE_
{mn,bc} 
+a_2^{(k)} \cE_{ij, ab}\cE_{kl,ac}\cE_{mn,bd}
\frac{t_{c}t_{d}}{t^2} \\
\nonumber
&+&
a_3^{(k)}(\cE_{kl, ab}\cE_{mn,bc}\cE_{mn,ad}+
(kl)\to (mn) )\frac{t_{c}t_{d}}{t^2}
+a_4^{(k)}\l \cE_{kl, ij}
\l \frac{t_mt_n}{t^2}-\frac{1}{d}\d_{mn}\r +(kl)\to (mn)  \r  \\
\nonumber
&+&
a_5^{(k)} \cE_{kl,mn}\l \frac{t_it_j}{t^2}-\frac{1}{d}\d_{ij}\r  
+
a_6^{(k)} 
\l \cE_{kl,ab}\cE_{ij,ac}
\l \frac{t_mt_n}{t^2}-\frac{1}{d}\d_{mn}\r + (kl)\to (mn)\r
\frac{t_{b}t_{c}}{t^2}  \\
\nonumber
&+&
a_7^
{(k)} \cE_{kl,ab}\cE_{mn,ac}\l
\frac{t_it_j}{t^2}-\frac{1}{d}\d_{ij}\r \frac{t_{b}t_{c}}{t^2}  
+a_8^{(k)} \l \frac{t_it_j}{t^2}-\frac{1}{d}\d_{ij}\r
 \l \frac{t_kt_l}{t^2}-\frac{1}{d}\d_{kl}\r
 \l \frac{t_mt_n}{t^2}-\frac{1}{d}\d_{mn}
 \r  \left.\right]
\eea
but with different coefficients $a_i^{(k)}$, where 
$i=1,\ldots,8$ and $k=1,2,3$.

For the first integral $I^1$ the coefficients 
$a_i^{(k)}$ are found to be
\bea
\la{a1}
\begin{array}{lll}  
a_1^{(1)} &=&
\frac{4(d-1)(d+1)}{d}+\frac{(d
-1)(8d+12)}{d^2} +\frac{8d+4}{d^2}+
\frac{12(d+2)+8(d-2)(d+3)}{d^2(d-1)(d+3)} 
\\
%%%%%%%%%%%%%%%
a_2^{(1)} & = &
-\frac{2(d-1)(2d+3)}{d}
-\frac{8d+4}{d} 
-\frac{12(d+2)+8(d-2)(d+3)}{d(d-1)(d+3)}
\\ 
%%%%%%%%%%%%%%%%%%
a_3^{(1)} & = &
-6
-\frac{2(d-1)(d+1)}{d}
-\frac{12(d+2)+8(d-2)(d+3)}{d(d-1)(d+3)} 
\\
a_4^{(1)} & = &
-\frac{(d+3)(d-3)+3(d+2)}{d(d-1)(d+3)}
\\
a_5^{(1)} & = &
-\frac{d-1}{d}
-\frac{3(d+2)+2(d-2)(d+3)}{d(d-1)(d+3)}
\\
a_6^{(1)} & = &
\frac{(d+2)(2d+1)}{d} 
+\frac{(d+2
)(6(d+2)+4(d-2)(d+3))}{d(d-1)(d+3)}
\\
a_7^{(1)} & = &
\frac{(d-1)(d+2)}{d} 
+\frac{(d+2)(6(d+2)+4(d-2)(d+3))}{d(d-1)(d+3)}
\\
a_8^{(1)}& =& -
\frac{(d+2)(d+4)(3(d+2)+2(d-2)(d+3))}{2d(d-1)(d+3)}.
\end{array}
\eea

For the second integral $I^2$ the 
coefficients $a_1^{(2)},\ldots a_8^{(2)}$ are as follows
\bea
\la{a2}
\begin{array}{ll}
a_1^{(2)}=\frac{2(d-1)}{d}+\frac{4}{d}+\frac{4(d+2)}{d(d-1)(d+3)}
&
a_2^{(2)}=-2(d-1)-4 -\frac{4(d+2)}{(d-1)(d+3)} \\
a_3^{(2)} =-2-\frac{4(d+2)}{(d-1)(d+3)
}
&
a_4^{(2)}=-1 -\frac{(d+2)}{(d-1)(d+3)} \\
a_5^{(2)} =-\frac{(d+2)}{(d-1)(d+3)}
&
a_6^{(2)}=d+2 +\frac{2(d+2)^2}{(d-1)(d+3)} \\
a_7^{(2)}=\frac{2(d+2)^2}{(d-1)(d+3)}
&
a_8^{(2)}=-\frac{(d+2)^2(d+4)}{2(d+3)(d-1)}
\end{array}
\eea
while for the third one they are 
\bea
\la{a3} 
\begin{array}{ll}   
a_1^{(3)}=-\frac{4}{d}-\frac{4(d+4)}{d(d-1)(d+3)}
&
a_2^{(3)}=4 +\frac{4(d+4)}{(d-1)(d+3)} \\
a_3^{(3)} =2+\frac{4(d+4)}{(d-1)(d+3)}
&
a_4^{(3)}=d+1 +\frac{(d+4)}{(d-1)(d+3)} \\
a_5^{(3)} =\frac{(d+4)}{(d-1)(d+3)}
&
a_6^{(3)}=-d-2 -\frac{2(d+2)(d+4)}{(d-1)(d+3)} \\
a_7^{(3)}=-\frac{2(d+2)(d+4)}{(d-1)(d+3)}
&
a_8^{(3)}=\frac{(d+2)(d+4)^2}{2(d-1)(d+3)}.
\end{array}
\eea

The coefficients $a_i^{(k)}$ are obtained for the 
general dimension $d$. We then specify their value 
for $d=4$ in the Table 1.

\bea
\nonumber
\begin{array}{|c|c|c|c|c|c|c|c|c|}
\hline
a_i^{(k)} & a_1^{(k)} & a_2^{(k)} & a_3^{(k)} & a_4^{(k)} & a_5^{(k)} & 
a_6^{(k)} & a_7^{(k)} & a_8^{(k)}  \\ \hline
k=1    
 & 547/21 & -1163/42 & -659/42 & -25/84 &  -109/84 & 843/42 &  465/42 & -92/7  \\
\hline
k=2     & 39/14  & -78/7    &  -22/7  &  -9/7  &  -2/7    & 66/7   & 24/7   
& -48/7    \\ \hline
k=3     & -29/21 & 116/21   & 74/21   & 113/21 &  8/21    & -74/21  & 
-32/21 &  64/7    \\ \hline
\end{array}
\eea
\begin{center}
Table 1. Coefficients $a_i^{(k)}$ for $d=4$.
\end{center}

\newpage
%%%%%%%%%%%%%%%%%%%%%%%%%%%%%%%%%%%%%%%%%%%%%%%%%%%%%%%%%%%%%%%%
 
\end{document}